# Causal Factors, Benefits and Challenges of Test-Driven Development: Practitioner Perceptions

Jim Buchan, Ling Li, Stephen G. MacDonell

*SERL, School of Computing and Mathematical Sciences*
*AUT University, Private Bag 92006*
*Auckland 1142, New Zealand*
jim.buchan@aut.ac.nz, angela.linz@gmail.com, stephen.macdonell@aut.ac.nz

**Abstract**

*This report describes the experiences of one organization's adoption of Test Driven Development (TDD) practices as part of a medium-term software project employing Extreme Programming as a methodology. Three years into this project the team's TDD experiences are compared with their non-TDD experiences on other ongoing projects. The perceptions of the benefits and challenges of using TDD in this context are gathered through five semi-structured interviews with key team members. Their experiences indicate that use of TDD has generally been positive and the reasons for this are explored to deepen the understanding of TDD practice and its effects on code quality, application quality and development productivity. Lessons learned are identified to aid others with the adoption and implementation of TDD practices, and some potential further research areas are suggested.*

**Keywords:** test-driven development (TDD), TDD benefits, TDD challenges, causal network

## I. INTRODUCTION

Test-driven development (TDD), which emphasizes a mind-set that functional code should be changed only in response to a failed test, is considered "proven practice" by many contemporary software development practitioners and text-book writers. Although it is a technique that has been practiced for decades [1], it has recently gained more visibility with the rise in use of Agile methodologies such as Extreme Programming (XP), where it is a core practice [2]. Proponents of TDD have reasoned that its use should result in improvements to code quality [3], testing quality [4], and application quality [5], compared to the traditional Test-Last (TL) approach. It has also been claimed to improve overall development productivity, encourage early understanding of the scope of requirements (user stories), as well as potentially leading to enhanced developer job satisfaction and confidence [3].

In contrast, critics claim that the frequent changes to tests in TDD are more likely (than in TL) to cause test breakages, leading to costly rework and loss of productivity [6]. Boehm and Turner [6] also note that with TDD the consequences of developers having inadequate testing skills may be amplified, compared to the consequences for a TL approach. Other critics note that TDD may not be appropriate for all application domains [7].

While these claims and criticisms provide some basis for evaluating the possible utility of TDD, practitioners and researchers have recognized the need for stronger evidence as a basis for investing – or not – in the effort to adopt and implement this set of practices. Recently, empirical researchers have been investigating the claimed benefits, constraints, and applicability of TDD in a variety of industrial and academic settings to build up a body of evidence. This empirical evidence is mixed in its results regarding the benefits of TDD (covered in more detail in Section VI).

This report adds further evidence regarding TDD in practice by describing the experiences of a software development team that has used TDD for three years in a specific project. This project (referred to as the "TDD project") adopted Extreme Programming (XP) as a development methodology. This was the first project to adopt TDD in this organization and the first experience of TDD for most of the team members (apart from the project leader) when the project began.

The aim of our study was to identify the benefits and challenges of using TDD in the experience of this project team. Further than this, the study aimed to deepen the understanding of these benefits and challenges by exploring what the team members perceived to be contributing TDD-related factors. The identified benefits and challenges and their perceived underlying causal factors add to the body of qualitative empirical evidence related to TDD practice in industrial settings.

We next provide a short description of the key aspects of TDD leading into some of the claimed benefits of the approach. The industry setting and data collection approach are described in Sections III and IV, followed by the presentation and discussion of our results in Section V. We then consider our results in the context of prior work and summarize the lessons learned in our study. We conclude our paper in Section VIII with consideration of further avenues for research.

## II. TDD BACKGROUND

TDD is a core practice in Extreme Programming (XP) involving Test-First (TF) development, closely intertwined with the complementary XP practices of automating testing, continuous testing and refactoring [8]. Any benefits and challenges attributed to any combination of these practices with a TF approach have been considered in this study.

There are several key characteristics of TDD described in the literature. Firstly there is the test orientation of TDD. TDD starts with thinking about how to test a small piece of selected functionality. The *first* coding task is the planning and writing of automated (unit) tests that would test if the functional requirement is met. Developers write a few tests for each small piece of functionality *before* starting the code for that functionality. These tests break the system and each test failure drives the activity of writing just enough production code to pass the failed test. This is in contrast to Test-Last development where tests are written *after* the target product features exist and are for verification and validation purposes. As emphasized by Lui and Chan [9], the tests in a TTD approach are used as the specification and scope for the functionality to be implemented, as well as being used for its verification and validation. TDD is therefore often described as a design technique rather than a testing technique since the test cases define what is required of each unit and this drives the design of the application [3].

Another key characteristic of TDD is the incremental and iterative nature of the process. Tests are added gradually during the development process. Typically the unit of testing is smaller than a user story, in contrast to TL development.

Figure 1 shows a clear comparison of the TF approach of TDD and the TL approach.

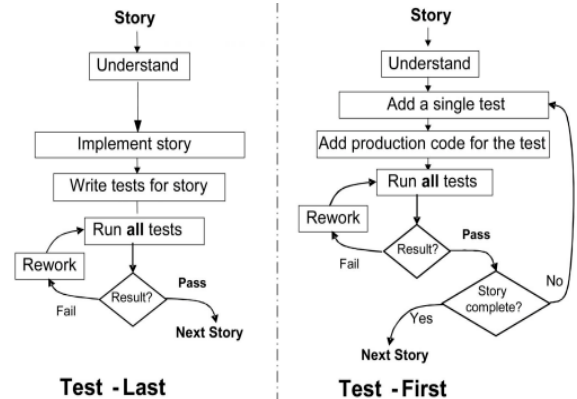

**Figure 1.** Test-first versus test-last development [10].

Using TDD, according to Beck [8], a developer:

1. quickly writes a new test for a small piece of functionality that is part of a user requirement (e.g. part of a user story)
2. runs *all* the tests to watch the new test fail
3. makes small changes to the production code (that are expected to pass the tests)
4. runs *all* tests to watch them all succeed (i.e. runs a regression test)
5. refactors the production and test code to improve their internal structure and remove redundancy if all tests pass (otherwise re-works the code until the tests are passed).
6. checks in both test and production source code at end of the day.

This cycle is repeated and the features and design of the product evolve incrementally. Typically steps 1-3 are done on a minute-by-minute time scale and steps 4 and 5 done periodically throughout the day.

Another notable aspect of TDD is the automation of the unit tests using frameworks such as xUnit. This means that the frequent regression testing that is an integral part of TDD is automated, and this set of automated tests is an increasingly important asset throughout the project and beyond, into maintenance.

As mentioned in the Introduction, there are a number of claimed beneficial outcomes that are said to accrue as a result of adopting TDD practices correctly. Within the TDD literature these benefits are generally implicated as contributing to one or more of the following high level gains: (1) improved code and design quality, implying that the code is easier to change and maintain; (2) enhanced application quality, resulting in a (more) reliable application that delivers the expected value to the client; and (3) better productivity, resulting in cost- and time-effective development. The intention of this study was to categorize the perceived benefits and challenges and causes factors identified by practitioners using TDD as contributing to one or more of these three high level benefits, enabling us to link the results of this study to prior TDD literature.

## III. INDUSTRIAL SETTING

The organization in focus here, based in Auckland, New Zealand, had spent around 15 years developing a software product, mainly in a 4GL development environment. A traditional Waterfall software development process, involving a Test-Last approach in particular, was used by the organisation. The ongoing maintenance, extension and customization of this legacy code to meet the needs of new markets and customers was becoming increasingly challenging. The decision was made to re-write the code in Java using a more modern technology stack with the specific aim of addressing this challenge. Management decided to adopt an Agile software development approach for this new project. Since there was little existing expertise in this area within the organization, a new project leader was hired, with specific experience in implementing XP practices. This project leader became the champion of XP, and in particular TDD, within this project team, based largely on his previous positive experiences using XP.

The re-write project was expected to take several years, involving millions of lines of legacy code. The project was in its third year at the time of this study. The size of the project team varied from 12 to 35 people at different times throughout the project, with staff at a range of experience levels including junior, intermediate and senior team members. The project team was split into several development teams typically comprising 5-6 members including a team leader. Pair programming was also practiced in the development teams, and so TDD was often undertaken in pairs. Some team members were moved into and out of the project team to work on other projects involving maintaining or extending the existing (legacy) product, as the need arose.

Most of the team members had little or no experience with TDD in practice prior to their involvement with this project and none had any formal training in TDD. Many of the project team members were initially skeptical about the benefits of TDD and so at first TDD was implemented with varying degrees of commitment.

A Business Analyst (BA) acted as a client proxy for the team and provided clarification of requirements to the development teams. Eclipse was used as the integrated development environment for all coding and jUnit for automated unit testing. Selenium was used to automate some aspects of acceptance testing.

## IV. DATA COLLECTION APPROACH

Experiences and views regarding the use of TDD compared to a non-TDD approach were gathered by interviewing five key team members who had been heavily involved in the TDD project. The five participants were interviewed separately using a semi-structured interview approach, and each interview lasted between 1.5 and 2 hours. Extensive interview notes were taken by two of the authors of this paper and the interviews were all audio recorded for transcription and later reference.

The five interviewees had all been involved in both Test-Last and Test-First practices within the company and so had a basis on which to compare and contrast the two approaches. The participants had between 7-15 years of experience related to software development and so were all quite experienced in their development roles. The interviewees all participated in day-to-day development work in the TDD project, four in a team leader role and one in a business analyst role. They contributed regularly to TDD-based coding and testing tasks, as well as having these key roles in the project team. These individuals were selected for interview because they not only had *personal* experience with TDD and non-TDD approaches in projects but also they had a wider appreciation of the views of the other team members because of their leadership role. In addition the nature of their roles made it more likely that they would be aware of the high level quality and productivity dynamics.

## V. RESULTS

This section summarizes the main results obtained from analyzing the interviews. It is structured to highlight the benefits and challenges of using TDD and their causes, along the dimensions of the high level benefit categories of code quality, application quality, and developer productivity.

### A. Positive Perceptions

The experience in using a TDD approach as part of their software development process was reported by all five interviewees as generally positive to them personally. They also observed that the view of TDD throughout the development teams had moved from skepticism at the start of the project to general support as providing tangible benefits. One interviewee noted that there was never strong criticism of TDD practices during team retrospectives as evidence of this.

When asked what they thought the *main* benefit of TDD was the participants' answers included:

- *throughout the development process the engineers think more widely*
- *higher quality code-it does what it is supposed to do*
- *increased test coverage*
- *confidence in the quality of code delivered to the client*
- *more client and developer confidence that the code will do what we said it will do*

These high level benefits show an emphasis on perceiving improvements to code and application quality, resulting in increased confidence in the application for developers and clients.

When questioned further about benefits, the interviewees identified a range of what they believed to be beneficial outcomes related specifically to TDD practices. These are summarized in the next three sub-sections. The interviewees attributed these benefits to changes in practice, attitude and behavior as a result of following TDD, compared to their non-TDD experiences.

### 1) Code Quality

Participants were unanimous in their perception that the use of TDD improved the quality of code compared to their experiences with traditional TL software development. They claimed that the use of TDD encouraged the development of simple, clean and meaningful code, more so than TL encouraged these outcomes. There was the perception that the discipline of following TDD would naturally develop habits that lead to better code as part of developers' everyday practice. The implication was that it was easier to be "lazy" and get away with developing messy or untested code using a TL approach. As one interviewee put it:

> *TDD helps developers towards simple designs; keeps things typically OO [Object Oriented] structured; pushes developers towards separated components.*

TDD was also viewed as guarding against the pressure of management to deliver code more quickly and compromising code quality, since this was not an option with TDD.

Participants identified several underlying factors that are consequences of following TDD practice, which they viewed as strongly contributing to the improvement of code quality. Interviewees perceived that a deeper understanding of the functionality required of a piece of code was a consequence of writing tests first. They stated that often, while trying to write a test for a piece of specified functionality (as part of a user story), they uncovered uncertainty in the meaning or scope of that functionality and sought clarification from the (proxy) client.

They viewed this benefit of test writing as being amplified when the test is written *before* the functional code because:

(1) better understanding triggered by the test-writing did not have the potential to require changes to the associated functional code, as it did with a TL approach; (2) there was no opportunity to defer (or omit) writing the test, in contrast to TL; (3) the extra effort invested into understanding what the functional code had to do *before* writing it, compared to TL, more often resulted in a clear idea of the objects and methods required in the functional code, making it easier to avoid cluttered code and have a simple design; (4) the tests and functional units tended to be smaller in scope than those written using a TL approach, making it easier (cognitively) to simplify the design, with fewer factors in the code to consider.

Another major contributor to good code design that interviewees identified with TDD was developers' increased confidence and willingness to put effort into improving the design of "perfectly good working code" through refactoring. This benefit was attributed to two main outcomes of TDD that are less "front of mind" in TL development: (1) refactoring is part of "how we do things" in TDD but was not emphasized to the same degree in the participants' TL experiences (2) the set of automated tests, also inherent to TDD, increased developer confidence to refactor code or try out new ideas because they could immediately run a set of automated tests to see if they "broke" existing code. They contrasted this with the "if it isn't broke, don't touch it" mentality they tended to have with a TL approach. They claimed that this increased confidence to change code, together with the fact that the "chunks" of code tended to be smaller using TDD, increased their willingness to refactor the code to improve design *during the implementation process*, resulting in better code design compared to using TL practices.

Participants also described the code written while using TDD practices as likely to be more "readable" and easier to maintain, compared to code written using TL. They saw this as also contributing to code quality. They explained that in writing the tests first they tended to use meaningful test, variable and class names, and that the use of meaningful names naturally carried over to the production code. Interviewees indicated that the associated tests also contributed to better understanding of code functionality. The interviewees observed that readability was further enhanced because fewer comments in the code were needed, since much of the information and semantics traditionally captured – or not – in comments were embodied in the associated tests. They also noted that the comments that *were* still required were more likely to be included (and were more likely to be useful), because so few were required. One interviewee makes this point as the following:

> *There is no need to have a large bit of documentation outside of the code or inside it to describe the functionality itself, as the test class would give a good idea of what the relevant class should be doing.*

*2) Application Quality*

Increased software reliability, as indicated by fewer defects in the release of an application, was attributed to higher quality testing during implementation. According to the interviewees, one result of writing breakpoint tests *before* production code was that developers spent more time designing the boundary cases needing to be covered by these tests, compared to when they were using the TL approach. This was described as resulting in more thorough testing and fewer defects at the end of the development cycle.

Higher test density and test coverage were also perceived to be encouraged by the use of TDD practices. Participants noted that with TL development there was a higher likelihood that testing would be left out or restricted to critical functionality, particularly if time was short. They noted that it was quite rare to go back and write unit tests for production code that had no associated tests, particularly if it had no errors associated with its functionality during acceptance testing or in use by the client. TDD instituted the discipline of *all* functionality being associated with a set of automated unit tests. This resulted in more tests and higher test coverage of the code. This improved quality of testing was perceived as being closely associated with improved application quality.

Another TDD practice that interviewees perceived as contributing to fewer errors propagating to the acceptance testing and final production code was the ability to run all automated tests before and after new functionality was implemented. The immediacy of the feedback from this continuous regression testing meant that appropriate rework of the production or test code was undertaken before developing additional functionality. This avoided compounding errors or inappropriate tests, and enabled developers to address issues in their context while still fresh in their mind.

Interviewees noted that the team appeared more confident handing over a new release of the software to the client compared to when they were using TL techniques. They observed that the tension previously associated with uncertainty around software releases seemed to be reduced. Increased developer confidence in the code reliability during and after development was a strong recurring theme in the interviews.

As noted in the subsection addressing code quality, participants reported that TDD encouraged better understanding of requirements by promoting more time spent early in development analyzing scenarios and business requirements and more frequent contact with the client or BA to discuss requirements and user stories. One interviewee described this in the following way:

> *Certainly, when you write a test it makes you question. You have to think about certain scenarios a little bit more. That makes you question your functionality that you are developing the code for, and then the questioning make you understand the requirement better.*

The interviewees saw this early, deep requirements understanding encouraged by TDD as leading to a product release that had less unwanted or incorrect functionality, as a result of fewer misinterpretations of the client's needs. This led to the team more frequently achieving higher levels of customer satisfaction.

*3) Productivity*

The five interviewees were categorical in their perception that the use of TDD had improved overall development productivity compared to their previous experiences with the TL approach. They explained that while the use of TDD may increase the time spent on developing tests and production code *during early phases*, as the development progressed, adding and testing new functionality was quicker and required less rework, compared to TL. They detailed a number of factors as contributing to improved productivity which they considered closely related to TDD practices.

It was perceived by the participants that under TDD more errors or misunderstandings were identified and addressed early in the development lifecycle compared to the TL approach. As one interviewee put it:

> *It's a lot cheaper in terms of resources to fix the issue immediately rather than months down the track when they may be discovered.*

They observed that re-work earlier in development was less time-consuming than re-work due to errors picked up later in the development lifecycle. The errors tended to be picked up earlier as a consequence of the higher quality testing and early deep requirements understanding, consequences of TDD that were also discussed in relation to application quality. In addition, the enhanced understanding often resulted in less functional code being written, saving time. This was a result of clarity of functional scope and writing only sufficient code to pass the associated tests.

Interviewees associated the use of TDD with improved understandability of the code, and they saw this as another factor contributing to improved productivity, particularly during application maintenance. Developers reported that this resulted in their being able to understand the implemented functionality more quickly, particularly when working on code that had been written by others or which had not been worked on for a long period of time.

Participants also described an increase in their satisfaction and motivation with TDD, resulting in improved personal productivity. The implication from this increased motivation was that less time was spent "off task" and coding tasks were done more quickly because of this improved attitude.

Enhanced motivation during implementation was a result of the improved confidence in their code working (passing tests), improved confidence that the code would do what the client wanted, a higher willingness to change code, improved understanding of the existing code, and early "success" when code passed a test. These TDD-related factors have all been discussed in previous sections, and are identified here by the participants as contributing holistically to a higher level of satisfaction and motivation, and consequent productivity gains. One interviewee expressed this view as:

> *It is frustrating to write 15 tests but it's more frustrating to not be able to change something or not to know your changes are safe.*

In addition some interviewees noted that with TDD developers were able to start writing (test) code sooner and "guilt-free", compared to a TL approach, where larger pieces of functionality tended to be analyzed and understood before beginning coding. The point being made was that developers like to write code and so this was a motivating factor. Finally, interviewees expressed the perception that more tests were re-used using TDD compared to their non-TDD practices, resulting in potential gains in ongoing productivity.

### *B. Perceived Challenges*

While the overall perception of the five key personnel was that TDD provided a number of benefits in practice, the interviewees also identified a number of challenges to adopting and implementing TDD.

Although it was noted that the large majority of developers could see the benefits of using TDD after a period of time, one interviewee noted that there was a constant need to (re)convince a subset of the project team of the benefits. For some team members doing TDD never became "natural" and they required reminders of the benefits of their effort.

In addition, interviewees observed that there is a large overhead in learning how to implement TDD practices, particularly if a developer has many years of non-TDD experience. They noted that a critical success factor in their adoption of TDD practices for the project being studied was the fact that the project leader was very convincing in his description of the benefits of TDD and his own positive experiences. Some participants felt that the project team took close to 1 year of using TDD to integrate and internalize TDD and visibly realize the improvements in code quality and productivity identified in this study. Participants noticed that new team members could still take several months to "change their mindset" and use TDD effectively. One participant viewed the introduction of new team members with weak TDD skills as particularly disruptive to teams, sometimes causing conflict and lowering morale.

Four of the interviewees expressed the view that realization of the benefits of using TDD is strongly linked to developers' capabilities in refactoring and writing high quality tests. They saw the development of these skills as one of the biggest challenges of TDD. They commented that more formal training in these skills, as well as the TDD process, would have been beneficial to their productivity.

Misunderstanding of TDD by upper management was viewed as a challenge by some interviewees. They explained that upper management often saw developers spending long periods of time on test writing rather than "getting on with the functional code". They interpreted this, often correctly, as spending longer time than before (using non-TDD methods) on providing functionality. The benefits to downstream development were often not so apparent.

One interviewee had the perception that TDD sometimes "gets in the way of simple, well-known code". It was described as a challenge to find the motivation to go back and write tests for code that was being re-used and that had been used in production and was well tested, proven and well understood.

## VI. DISCUSSION

Although not without its challenges, overall, the interviewees described the change to using TDD as a positive move compared to their previous and ongoing experiences with traditional development methods. They perceived general improvements to application quality, code quality, overall productivity and personal satisfaction via a collection of contributing factors. Are these perceptions supported by others using TDD practices? Over the last decade a number of researchers have undertaken empirical studies related to the effectiveness of TDD and the remainder of this section considers these as a basis for comparison with the interviewees' perceptions.

A number of studies have investigated the quality of applications developed using TDD compared to using TL. Results are varied and inconclusive, however. One early study [11] involved a controlled experiment with 19 students. The students were divided into two groups and both teams were tasked with developing a small Java program, one using a test-first approach and the other a traditional test-after-coding approach. The quality of each final application was measured in terms of its reliability, in this case by the number of acceptance-tests failed. The result was that the group using TF did not produce a more reliable application. Furthermore, they took around the same time to produce this poorer quality application, despite showing signs of better program understanding. Similar controlled experiments [12-15] do not show any benefit to application quality in using TDD compared to using various TL techniques. On the other hand, other controlled experiments [5, 16-17] have indicated that there may be a positive relationship between the use of

TDD and application quality. In [16] an experiment was conducted with 24 professionals and the final applications developed using TDD passed 18-50% more external tests than the applications developed with TL. Other investigations involving case studies of industrial projects [18-21] consistently report that TDD provided significant improvements to application quality. It is interesting to note that it is the studies of industrial projects "in the wild" that provide the most consistent findings regarding application quality improvement with TDD. The industrial study reported in this paper aligns with these findings and adds to this body of evidence. This study also suggests a number of "causal pathways" of factors that may contribute to application quality. It would be useful to investigate the strength of these relationships in other industrial project settings to further deepen the understanding of the effects of TDD on application quality.

The effects on code quality of using TDD have also been investigated. In one controlled case study [22] teams of students undertook four different software projects for real customers, two using TDD and two using iterative TL approaches. The TDD teams produced higher code quality, as measured using a suite of well-defined traditional metrics for code quality (although the code quality was still quite high for the TL teams). A similar study of students doing projects for industry clients [17] showed that TDD teams tended to produce code with smaller, less complex classes than teams using TL (although cyclomatic complexity was very similar in both cases). Another experimental pilot study using students [23] is not conclusive but does indicate that the use of TDD seemed to result in higher design quality compared to using TL. In the post-experiment survey of the study done by [16] 79% of the participants perceived TDD as promoting simpler code design and 92% as resulting in higher quality code, compared to their use of TL. Other experiments [11, 13, 15, 24-25] show either no change or even a decrease in code quality with the use of TDD practices. The uncertainty in the conclusion to draw from these mixed results is in direct contrast to the apparent certainty of the study reported in this paper, where all five interviewees had very strong perceptions that the change to TDD had resulted in improved code quality. This suggests that further study of the effects of TDD on code quality in industrial projects would be a fruitful area for investigation, particularly given that other industrial case studies [18-21] did not consider code quality specifically.

Related to quality is the notion of testing quality and improvements to testing quality is an outcome of several experiments involving the use of TDD. For example in the study by [16], TDD use produced very high test coverage (98% code, 92% statement and 97% branch coverage). Improvements to test coverage, test effort, test volume, or testing frequency are also reported by [12, 14, 26]. These findings align well with the findings of this study where participants noted increased test density and coverage with TDD as a contributing factor of application reliability.

Several recent empirical studies also investigated TDD-related effects on aspects of productivity. In one controlled experiment employing 24 students [10], the group undertaking TDD practices show improved developer productivity, reduced debugging effort and reduced rework, compared to the control group using iterative TL development. Several other controlled experiments [17, 23-24] also support higher developer or overall productivity with TDD compared to TL approaches. There is no clear conclusion, however, since several other experimental studies [11, 15, 27] and the two industrial case studies that have considered productivity [18, 28] all suggest that the use of TDD either has no measurable effect on productivity or even reduces productivity by between 15% and 35%. Interestingly, in the post-experiment survey of the subjects in [16] 78% of the respondents indicated an improvement to overall productivity, 95.8% perceived a reduction in debugging time, and 50% felt there was a reduction in coding time. This is despite the TDD groups being measured as taking 16% longer to complete the assignment than the TL group. The perceptions of the interviewees in our study are closely aligned with this survey data [16]. The interviewees described improvements to overall productivity as being due to, among other factors, reduced debugging effort (since errors were caught and fixed early). It would have been useful also to have had access to quantitative measures of productivity for our study.

In light of the mixed results reported by previous empirical studies, the findings in our study generally support many of the positive findings reported, particularly in the industrial case studies. Moreover, this study deepens the understanding of this evidence by identifying influencing factors. This adds to the body of evidence with which practitioners can base the decision to use TDD or not and provides some guidance on areas to continue researching. It is only by building up such evidence over many projects that a clear picture of the benefits and constraints of TDD use will be formed.

## VII. LESSONS LEARNED

The interviewees identified a number of lessons and recommendations as a result of reflecting on their experiences on TDD within this project. The main points are summarized as follows.

*Each team leader should be a champion of TDD.* Members of development teams that had team leaders who were less committed to the practices of TDD tended to be inconsistent in their use of TDD in their code development. This sometimes resulted in untested code, incomplete test-suites, and tension within the team. It would also be of benefit if the team leader had sufficient previous experience with TDD that they could mentor other team members and recognize when TDD practices were not being followed.

*TDD should be followed strictly to realize the benefits.* There are a number of misconceptions and common mistakes in implementing TDD into daily development practice; it is more than just writing tests first! In [29], 9 common mistakes are self-identified by 218 TDD practitioners. Incorrect implementation of TDD may mean that the benefits evidenced and reasoned in the literature are not realized. It was felt that a common base level of TDD training and inter-team reviews of practice could help to ensure that TDD is practiced correctly. It was the experience of some teams that pair programming afforded a level of quality assurance on the TDD process also, provided the pairs had sufficient knowledge. Ensuring good access to the client or suitable proxy was also seen as important.

*Management should be made aware of the characteristics of software development using TDD.* This was not a lack of management "buy-in", but rather a lack of understanding of the implications of using a TDD approach. Management was supportive of using Agile methods and TDD *in principle*. But they were a source of tension when the coding of functionality took longer than expected based on management's previous non-TDD experiences.

*The benefits of TDD should be made more visible.* It was felt that the *costs* of TDD were more visible than the benefits generally, and this contributed to variable levels of commitment to TDD among team members, as well as pressure from upper management. At an individual level the outcomes of TDD accumulate over a period of time and may never be shared with the rest of the project team or get "forgotten" over time. Team leaders and project managers are in the best position to proactively capture and disseminate both qualitative and quantitative indications of the benefits of using TDD. Measures of the most valuable benefits related to TDD should be tracked and monitored and integrated into the daily process. These can then be displayed and shared among the project team members.

*Planned induction and training needs to be provided for developers new to TDD.* It was identified that new team members with low TDD experience could be disruptive. In addition, while they improved their understanding of TDD practices in an *ad hoc* manner with on-the-job training, they were generally below the average productivity of the team. To expedite this induction of new team members, a planned programme of both formal training and on-the-job training in applying TDD was suggested. It was noted that this programme should include training on writing high quality tests and refactoring.

*Training should be provided to up-skill developers in their test writing and refactoring capabilities.* Since these two skills are so critical to successful TDD in practice, it was felt that knowledge sharing and training in these areas should support continuous improvement and learning.

*Tools and techniques for supporting TDD should be integrated into the everyday development environment.* One of the challenges of implementing TDD identified by all interviewees was the lack of tool support for TDD practices provided by their development environment (Eclipse in this case). An integrated development environment (IDE) as well as other relevant specialized tools that provide TDD support could overcome this frustration and improve productivity.

Figure 2 summarizes the benefits of TDD as well as the causal factors that contribute to those benefits, as identified by the interviewees, in the form of an interconnected network. This network is based on an analysis of the interview field notes and transcripts into themes and concepts, and their causal relationships. This provides a concrete overview of our learning about the relationships between the quality and productivity benefits and contributing factors related to TDD practice. An important learning is the complex nature of the interactions of different factors that can influence productivity and quality outcomes. This implies that it cannot be reasonably expected that specific outcomes will be totally predictable over a wide range of organizational and project contexts. It also provides further motivation for studying the phenomenon in a realistic setting, where understanding the richness of these interactions can be used to interpret experiments and suggest new hypotheses to test.

## VIII. CONCLUSIONS AND FUTURE WORK

We set out to understand how TDD was adopted and implemented in an organization more familiar with TL approaches to development, with a particular focus on how and when TDD-related benefits could be achieved. Our analysis, based on interviews with five key personnel, highlights that benefits in terms of improved code quality, application quality and productivity can be achieved in time if certain factors are addressed and particular conditions put in place. This has enabled us to derive a network of factors that interact and together influenced the outcomes for those practicing TDD in our industrial partner organization. While it cannot be claimed that the proposed network of factors is generalisable to other projects, it does provide a starting point for future hypothesis generation and testing that could provide deeper understanding and add to the body of empirical evidence related to TDD.

Our analysis relies on an assumption that the organization was indeed *practicing* TDD correctly. The participants seemed to have internalized the main concepts of TDD as described in Section II, however their actual practices were not observed.

Also, it could be contended that the perceived benefits actually derived from the organization's use of XP and iterative incremental development, and not specifically from TDD. We therefore encourage future industry-based research studies that consider specific aspects of organizational processes, and that test the derived relationships and pathways through our causal network.

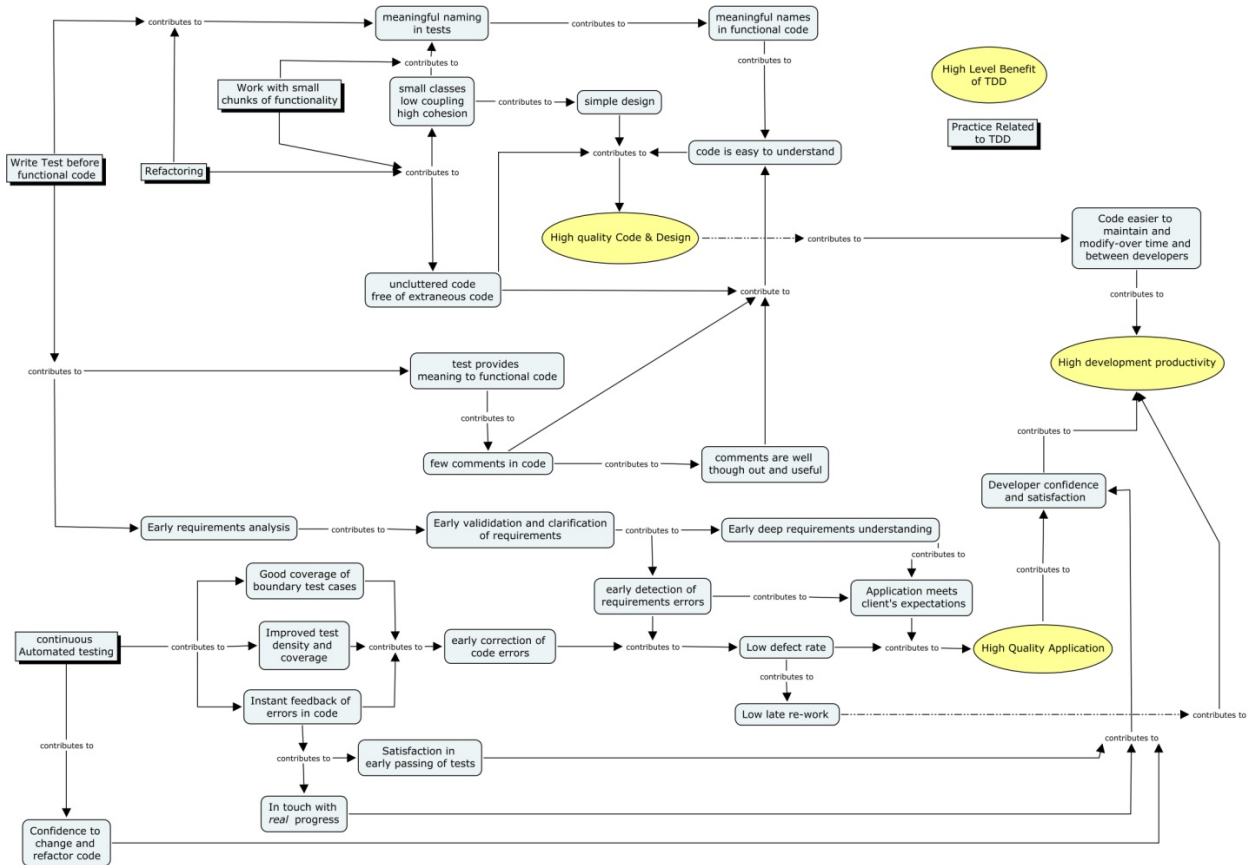

**Figure 2.** Causal network of TDD benefits and contributing factors.

## ACKNOWLEDGMENT

We thank our industrial partner and in particular the willing interviewees. Ling Li was employed by the partner organization when the TDD project was undertaken.

## REFERENCES

[1] C. Larman and V. R. Basili, "Iterative and incremental developments. a brief history," *Computer,* vol. 36, pp. 47-56, 2003.

[2] K. Beck, *Extreme Programming Explained: Embrace Change*. Reading, Massachusetts, USA: Addison Wesley, Longman, 2000.

[3] K. Beck, "Aim, fire [test-first coding]," *Software, IEEE,* vol. 18, pp. 87-89, 2001.

[4] Test-Driven Development: A Practical Guide, 2003.

[5] L. Crispin, "Driving Software Quality: How Test-Driven Development Impacts Software Quality," *Software, IEEE,* vol. 23, pp. 70-71, 2006.

[6] B. Boehm and R. Turner, *Balancing Agility and Discipline - A GUide to the Perplexed*. Boston: Addison-Wesley, 2004.

[7] M. Stephens and D. Rosenberg, *Extreme Programming Refactored: The Case Against XP*. Berkley: Apress, 2003.

[8] K. Beck, *Test Driven Development: By Example*. Reading, Massachusetts, USA: Addison Wesley, Longman, 2003.

[9] K. M. Lui and K. C. C. Chan, *Software Development Rhythms : Harmonizing Agile Practices for Synergy*. Hoboken, N.J: Wiley-Interscience, 2007.

[10] H. Erdogmus, et al., "On the effectiveness of the test-first approach to programming," *Software Engineering, IEEE Transactions on,* vol. 31, pp. 226-237, 2005.

[11] M. M. Muller and O. Hagner, "Experiment about test-first programming," *Software, IEE Proceedings -,* vol. 149, pp. 131-136, 2002.

[12] A. Geras, et al., "A prototype empirical evaluation of test driven development," in *Software Metrics, 2004. Proceedings. 10th International Symposium on*, 2004, pp. 405-416.

[13] A. Gupta and P. Jalote, "An Experimental Evaluation of the Effectiveness and Efficiency of the Test Driven Development," in *Empirical Software Engineering and Measurement, 2007. ESEM 2007. First International Symposium on*, 2007, pp. 285-294.

[14] L. Huang and M. Holcombe, "Empirical investigation towards the effectiveness of Test First programming," *Information and Software Technology,* vol. 51, pp. 182-194, 2008.

[15] M. Pancur, et al., "Towards empirical evaluation of test-driven development in a university environment," in *EUROCON 2003. Computer as a Tool. The IEEE Region 8*, 2003, pp. 83-86 vol.2.

[16] B. George and L. Williams, "A structured experiment of test-driven development," *Information and Software Technology,* vol. 46, pp. 337-342, 2004.

[17] J. H. Vu, et al., "Evaluating Test-Driven Development in an Industry-Sponsored Capstone Project," in *Information Technology: New Generations, 2009. ITNG '09. Sixth International Conference on*, 2009, pp. 229-234.

[18] T. Bhat and N. Nagappan, "Evaluating the efficacy of test-driven development: industrial case studies," presented at the Proceedings of the 2006 ACM/IEEE international symposium on Empirical software engineering, Rio de Janeiro, Brazil, 2006.

[19] K. M. Lui and K. C. C. Chan, "Test Driven Development and Software Process Improvement in China," in *Extreme Programming and Agile Processes in Software Engineering*. vol. 3092, J. Eckstein and H. Baumeister, Eds., ed: Springer Berlin / Heidelberg, 2004, pp. 219-222.

[20] E. M. Maximilien and L. Williams, "Assessing test-driven development at IBM," presented at the Proceedings of the 25th International Conference on Software Engineering, Portland, Oregon, 2003.

[21] L. Williams, et al., "Test-driven development as a defect-reduction practice," in *Software Reliability Engineering, 2003. ISSRE 2003. 14th International Symposium on*, 2003, pp. 34-45.

[22] M. Siniaalto and P. Abrahamsson, "Does Test-Driven Development Improve the Program Code? Alarming Results from a Comparative Case Study," in *Balancing Agility and Formalism in Software Engineering*, M. Bertrand, et al., Eds., ed: Springer-Verlag, 2008, pp. 143-156.

[23] R. Kaufmann and D. Janzen, "Implications of test-driven development: a pilot study," presented at the Companion of the 18th annual ACM SIGPLAN conference on Object-oriented programming, systems, languages, and applications, Anaheim, CA, USA, 2003.

[24] D. S. Janzen and H. Saiedian, "On the Influence of Test-Driven Development on Software Design," presented at the Proceedings of the 19th Conference on Software Engineering Education \& Training, 2006.

[25] M. Müller, "The Effect of Test-Driven Development on Program Code," in *Extreme Programming and Agile Processes in Software Engineering*. vol. 4044, P. Abrahamsson, et al., Eds., ed: Springer Berlin / Heidelberg, 2006, pp. 94-103.

[26] C. D. Thomson, et al., "What Makes Testing Work: Nine Case Studies of Software Development Teams," in *Testing: Academic and Industrial Conference - Practice and Research Techniques, 2009. TAIC PART '09.*, 2009, pp. 167-175.

[27] G. Canfora, et al., "Productivity of Test Driven Development: A Controlled Experiment with Professionals," in *Product-Focused Software Process Improvement*. vol. 4034, J. Münch and M. Vierimaa, Eds., ed: Springer Berlin / Heidelberg, 2006, pp. 383-388.

[28] L.-O. Damm, et al., "Introducing Test Automation and Test-Driven Development: An Experience Report," *Electronic Notes in Theoretical Computer Science,* vol. 116, pp. 3-15, 2005.

[29] M. F. Aniche and M. A. Gerosa, "Most Common Mistakes in Test-Driven Development Practice: Results from an Online Survey with Developers," in *Software Testing, Verification, and Validation Workshops (ICSTW), 2010 Third International Conference on*, 2010, pp. 469-478.